\documentclass{article}
\usepackage{graphicx}
\begin{document}

\title{The effects of minimal length and maximal momentum on the transition rate of ultra cold neutrons in gravitational field}
\author{Pouria Pedram$^a$\thanks{pouria.pedram@gmail.com}, Kourosh Nozari$^b$\thanks{knozari@umz.ac.ir}, and S. H. Taheri$^b$\\
{\small $^a$Plasma Physics Research Center, Science and Research
Branch, Islamic Azad University, Tehran,
            Iran}\\{\small $^b$Department of Physics, Islamic Azad University, Sari Branch, Sari, Iran}}

\date{\today}

\maketitle \baselineskip 24pt

\begin{abstract}
The existence of a minimum observable length and/or a maximum
observable momentum is in agreement with various candidates of
quantum gravity such as string theory, loop quantum gravity,
doubly special relativity and black hole physics. In this
scenario, the Heisenberg uncertainty principle is changed to the
so-called Generalized (Gravitational) Uncertainty Principle (GUP)
which results in modification of all Hamiltonians in quantum
mechanics. In this paper, following a recently proposed GUP which
is consistent with quantum gravity theories, we study the quantum
mechanical systems in the presence of both a minimum length and a
maximum momentum. The generalized Hamiltonian contains two
additional terms which are proportional to $\alpha p^3$ and
$\alpha^2 p^4$ where $\alpha \sim 1/M_{Pl}c$ is the GUP parameter.
For the case of a quantum bouncer, we solve the generalized
Schr\"odinger equation in the momentum space and find the modified
energy eigenvalues and eigenfunctions up to the second-order in
GUP parameter. The effects of the GUP on the transition rate of
ultra cold neutrons in gravitational spectrometers are discussed
finally.
\end{abstract}


\section{Introduction}\label{sec1}
The modification of classical notion of the spacetime is one of
the common features of all quantum gravity theories. In these
theories, it is assumed that the usual concept of continuity of
the spacetime manifold would break down when we probe distances
smaller than the Planck length or energies larger than the Planck
energy. If this fact is confirmed by future experiments, it could
make a deep influence on our understanding about our surrounding
universe. On the other hand, it may help us to find the answer of
many unsolved problems such as the mechanism of singularity
avoidance at early universe and also the black hole spacetime.

One of the common properties of various candidates of quantum
gravity such as string theory, loop quantum gravity and doubly
special relativity is the existence of a minimum measurable length.
Also, some evidence from black hole physics assert that a minimal
length of the order of the Planck length arises naturally from any
theory of quantum gravity. In addition, in the context of
non-commutativity of the spacetime manifold, we also realize the
existence of a minimal measurable length.

Evidently, this assumption is in apparent contradiction with the
Heisenberg uncertainty principle which in principle agrees with
the measurement of highly accurate results for a particles'
positions or momenta, separately. In fact, in the Heisenberg
picture, the minimum observable length is actually zero. So, if we
are interested in to incorporate the idea of minimal length, we
need to modify the Heisenberg uncertainty principle to the
so-called Generalized Uncertainty Principle (GUP)
\cite{1,11,12,13,14,15,16,17,18,19,10,5,51,52,53,54}. In other
words, we should modify the commutation relations between position
and momentum operators in the Hilbert space (deformed Heisenberg
algebra). Moreover, in doubly special relativity theories, in
order to preserve velocity of light and the Planck energy as two
invariant quantities, the existence of a maximal momentum is
essentially required \cite{2,21,3}.

In GUP formalism, the idea of a minimum observable length and a
maximum observable momentum changes the usual form of all
Hamiltonians in quantum mechanics (see ref.~\cite{6} and references
therein). In fact, the modified Hamiltonians contain additional
terms proportional to the powers greater than two of the momentum.
So, in the quantum domain, the corresponding generalized
Schr\"odinger equation has a completely different differential
structure. More precisely, when we solve a forth-order generalized
Schr\"odinger equation in the position space, some solutions are
unphysical which should be discarded. However, if possible, it is
more desirable to reduce the order of the differential equation by
some methods such as solving the differential equation in the
momentum space.

In this paper, we consider a recently proposed GUP which is
consistent with string theory, doubly special relativity and black
hole physics and predicts both a minimum measurable length and a
maximum measurable momentum \cite{main,main2,p1}. For this purpose,
first we find the modified Hamiltonian of a general quantum
mechanical system up to the second order of the GUP parameter
$\alpha$. Then, for the case of a particle which is bouncing
elastically and vertically above a mirror in the Earth's
gravitational field we solve the generalized Schr\"odinger equation
in the momentum space and find the corresponding energy eigenvalues
and eigenfunctions up to ${\cal{O}}(\alpha^2)$. In particular, we
show that the existence of a maximal momentum reduces the effect of
a minimal length on the energy spectrum which results in the
reduction of the transition rate of ultra cold neutrons in
gravitational spectrometers with respect to the case that the
assumption of the maximal momentum is absent \cite{NP}.

\section{A generalized uncertainty principle}\label{sec2}
Recently, a GUP is proposed by Ali \textit{et al.} which is
consistent with the existence of the minimal measurable length and
the maximal measurable momentum \cite{main,main2}. In this proposal, the
spaces of position and momentum are assumed to be commutative
separately i.e. $[X_i,X_j]=[P_i,P_j]=0$. Also, the following
deformed Heisenberg algebra are satisfied
\begin{eqnarray}\label{xp}
[X_i, P_j] &=& i \hbar\left[ \delta_{ij}- \alpha \left( P\delta_{ij}
+ \frac{P_i P_j}{P} \right) + \alpha^2 \left( P^2 \delta_{ij} +
3P_{i} P_{j} \right) \right],
\end{eqnarray}
where $\alpha = {\alpha_0}/{M_{Pl}c} = {\alpha_0
\ell_{Pl}}/{\hbar}$, $P^{2} = \sum\limits_{j=1}^{3}P_{j}P_{j}$,
$M_{Pl}$ is the Planck mass, $\ell_{Pl}$ is the Planck length
$\approx 10^{-35}m$, and $M_{Pl} c^2$ is the Planck energy
$\approx 10^{19}$GeV. Using the above commutation relations, we
can obtain the generalized uncertainty relation in one-dimension
up to the second order of the GUP parameter \cite{main,main2}
\begin{eqnarray}\nonumber
 \Delta X \Delta P &\geq& \frac{\hbar}{2}
\left[ 1 - 2\alpha \langle P\rangle + 4\alpha^2 \langle P^2\rangle
\right],  \\  &\geq& \frac{\hbar}{2} \left[ 1 + \left(
\frac{\alpha}{\sqrt{\langle P^2 \rangle}} +4\alpha^2  \right)
\Delta P^2  +   4\alpha^2 \langle P \rangle^2
 -   2\alpha \sqrt{\langle P^2 \rangle}
\right].
\end{eqnarray}
The above inequality implies both a minimum length and a maximum
momentum at the same time, namely \cite{main,main2}
\begin{eqnarray}\label{Lmin}
\left\{
\begin{array}{ll}
\Delta X \geq (\Delta X)_{min}  \approx \alpha_0\ell_{Pl},  \\\\
\Delta P \leq (\Delta P)_{max} \approx \frac{\displaystyle
M_{Pl}c}{\displaystyle\alpha_0}.
\end{array}
\right.
\end{eqnarray}
We can also rewrite the position and momentum operators in terms of
new variables
\begin{eqnarray}\label{x0p0}
\left\{
\begin{array}{ll}
X_i = x_{i},\\\\ P_i = p_{i} \left( 1 - \alpha p + 2\alpha^2 p^2
\right),
\end{array}
\right.
\end{eqnarray}
where $x_{i}$ and $p_{i}$ obey the usual commutation relations
$[x_{i},p_{j}]=i\hbar\delta_{ij}$. It is straightforward to check
that with this definition, eq.~(\ref{xp}) is satisfied up to
${\cal{O}}(\alpha^2)$. Therefore, we can interpret $p_i$ and $P_i$
as follows: $p_{i}$ is the momentum operator at low energies
($p_{i}=-i\hbar
\partial/\partial{x_{i}}$) and $P_{i}$ is the momentum operator at high
energies. Moreover, $p$ is the magnitude of the $p_{i}$ vector
($p^{2}=\sum\limits_{j=1}^{3}p_{j}p_{j}$). To study
the effects of this kind of GUP on the quantum mechanical systems,
let us consider the following general Hamiltonian
\begin{eqnarray}
H=\frac{P^2}{2m} + V(\vec R).
\end{eqnarray}
Now, if we write the high energy momentum in terms of low energy one
(\ref{x0p0}), we obtain
\begin{eqnarray}\label{hamil}
H=H_0+\alpha H_1+\alpha^2 H_2+{\cal{O}}(\alpha^3),
\end{eqnarray}
where $H_0=\frac{\displaystyle p^2}{\displaystyle2m} + V(\vec R)$
and
\begin{eqnarray}
H_1=-\frac{p^3}{m}, \hspace{1cm} H_2=\frac{5p^4}{m}.
\end{eqnarray}
Therefore, in the GUP scenario two additional terms proportional to
$\alpha p^3$ and $\alpha^2 p^4$ appear in the modified version of
the Hamiltonian which the later is the result of the minimum length
assumption and the former is the result of the maximum momentum
assumption. In the next section, we consider the problem of a
quantum bouncer in GUP formalism and find its modified
eigenfunctions and eigenvalues up to ${\cal{O}}(\alpha^2)$ and
compare our results with the ones which the second assumption is
absent \cite{NP}. As an application, we show that GUP will affect
the transition rate of ultra cold neutrons bouncing above a mirror
in the Earth's gravitational field.

\section{Modification of a quantum bouncer's spectrum in GUP scenario}
\begin{figure}
\centering
\includegraphics[width=8cm]{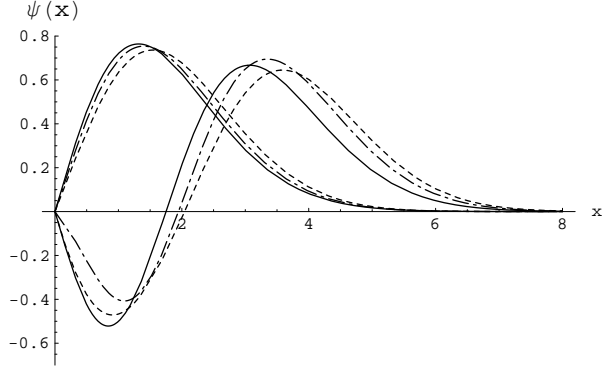}
\caption{The normalized ground state and first excited state
eigenfunctions of a quantum bouncer in the framework of the
generalized commutation relation (\ref{xp}) for $H_0$ (solid line),
$H_0+H_2$ (dashed line), $H_0+H_1+H_2$ (dot-dashed line), with
$\hbar=1$, $g=2$, $m=1/2$, and $\alpha=0.1$.} \label{fig1}
\end{figure}

To study the effects of GUP on the spectrum of a quantum bouncer,
let us consider a particle of mass $m$ which is bouncing elastically
and vertically on an ideal reflecting floor in the earth's
gravitational field so that
\begin{eqnarray}
V(X)=\left\{
\begin{array}{cc}
mgX&\hspace{2cm} X>0,\\\\ \infty\, &\hspace{2cm} X\leq 0,
\end{array}
\right.
\end{eqnarray}
where $g$ is the acceleration caused by the gravitational attraction
of the Earth. The Hamiltonian of the system is
\begin{equation}
H=\frac{P^2}{2m}+mgX,
\end{equation}
which using eq.~(\ref{hamil}) casts in the form of the following
generalized Schr\"odinger equation
\begin{eqnarray}
-\frac{\hbar^2}{2m}\frac{\partial^2\psi(x)}{\partial
x^2}-i\alpha\frac{\hbar^{3}}{m}\frac{\partial^{3}\psi(x)}{\partial
x^{3}}+5\alpha^2\frac{\hbar^{4}}{m}\frac{\partial^{4}\psi(x)}{\partial
x^{4}} +mgx\psi(x)=E\psi(x).
\end{eqnarray}
This equation is exactly solvable for $\alpha=0$ and the solutions
can be written in the form of the Airy functions. Also, the energy
eigenvalues correspond to the zeros of the Airy function. For the
case of $\alpha\ne0$, we encounter a quite different situation.
Because, the above equation is a forth-order differential equation
which in general admits four independent solutions. However, some of
these solutions are unphysical and should be discarded. One way to
obtain physical solutions is to reduce the order of the differential
equation which is fortunately possible in our case. In fact, if we
write the above equation in the momentum space, because of the
linear form of the potential term, it can be rewritten as a first
order differential equation. Since the first order equation is much
easier to handle, we define a new variable $z=x-\frac{E}{mg}$ and
rewrite the above equation in the momentum space
\begin{equation}
\frac{p^2}{2m}\phi(p)-\alpha\frac{p^3}{m}\phi(p)+5\alpha^2\frac{p^4}{m}\phi(p)+i\hbar
mg\phi'(p)=0,
\end{equation}
where $\phi(p)$ is the inverse Fourier transform of $\psi(z)$ and
the prime denotes the derivative with respect to $p$. It is
straightforward to check that this equation admits the following
solution
\begin{equation}
\phi(p)=\phi_0\exp\left[\frac{i}{6m^2g\hbar}\left(p^3-\frac{3}{2}\alpha
p^4+6\alpha^2 p^5\right)\right].
\end{equation}
Since $\alpha$ is a small quantity, we can expand the above solution
up to the second-order of $\alpha$ as
\begin{equation}
\phi(p)\simeq\phi_0\exp\left({\frac{ip^3}{6m^2g\hbar}}\right)\left(1+\frac{i
}{m^2g\hbar}\left[-\frac{\alpha
p^4}{4}+\alpha^2\left(p^5+\frac{ip^8}{32m^2g\hbar}\right)\right]+{\cal{O}}(\alpha^3)\right).
\end{equation}
Now, using the Fourier transform, we can obtain the solution in the
position space. Before writing the solution, we should be careful
about the nature of the terms appear in the above equation. Note
that, the terms which obey $\phi^*(p)=\pm\phi(-p)$ result in real
and imaginary terms in the wave function, respectively. So the first
term in the square brackets leads to an imaginary term in the wave
function. Since the Hamiltonian is hermitian, both the real and
imaginary parts of the wave function should satisfy the
Schr\"odinger equation and vanish at $x=0$, separately. Because the
unperturbed wave function is real and the condition
$\mbox{Im}[\psi(0)]=0$ does not depend on $\alpha$, the resulting
energy spectrum and consequently the imaginary wave function will
not be physical and should be discarded. This is in agreement with
the fact that bound states of one-dimensional quantum systems should
be real. Note that, the existence of an imaginary part of the wave
function is due to the presence of $p^3$ in the perturbed
Hamiltonian. We can also deduce this result by implementing the
perturbation analysis. It is straightforward to check that the
first-order correction of $H_1=-\alpha p^3/m$ to the wave function
is completely imaginary. However, since the unperturbed
eigenfunctions are real functions of $x$, the first-order correction
of $H_1$ to the energy spectrum is identically zero i.e. $\langle
n|\frac{-p^3}{m}|n\rangle=-\frac{i\hbar^3}{m}\int_{-\infty}^{\infty}(\psi_n^{0}(x))^*\frac{\partial^3}{\partial
x^3}\big(\psi_n^0(x)\big)\,dx\equiv0$. Putting these facts together,
we conclude that the effect of GUP on the eigenfunctions is at least
second-order in GUP parameter and up to a normalization factor we
have
\begin{eqnarray}
\mbox{Re}[\psi(x)]&=&\theta(x)\Bigg[\mbox{Ai}\left[\beta\left(x-\frac{E}{mg}\right)\right]+\alpha^2
m^2g\left(x-\frac{E}{mg}\right)\times\nonumber\\
&\times&\Bigg\{9\mbox{Ai}\left[\beta\left(x-\frac{E}{mg}\right)\right]
+\left(x-\frac{E}{mg}\right)\mbox{Ai}'\left[\beta\left(x-\frac{E}{mg}\right)\right]-\nonumber\\
&-&\frac{1}{4}\beta^3
\left(x-\frac{E}{mg}\right)^3\mbox{Ai}\left[\beta\left(x-\frac{E}{mg}\right)\right]\Bigg\}\Bigg],
\end{eqnarray}
where $\theta(x)$ is the Heaviside step function,
$\beta=\left(\frac{2m^2g}{\hbar^2}\right)^{1/3}$ and the prime
denotes derivative with respect to $x$. Finally, since the
potential is infinite for $x\le0$, we demand that the wave
function should vanish at $x=0$. This condition results in the
quantization of the particle's energy, namely
\begin{eqnarray}\label{zero}
&\mbox{Ai}&\left(-\frac{\beta E_n}{mg}\right)-\alpha^2
mE_n\left[9\mbox{Ai}\left(-\frac{\beta
E_n}{mg}\right)-\frac{E_n}{mg}\times\right.\nonumber\\
&\times&\left.\mbox{Ai}'\left[\beta\left(x-\frac{E_n}{mg}\right)\right]\Bigg|_{x=0}+\frac{E_n^3}{2mg^2\hbar^2}\mbox{Ai}\left(-\frac{\beta
E_n}{mg}\right)\right]=0.
\end{eqnarray}
To proceed further and for the sake of simplicity, let us work in
the units of $\hbar=1$, $g=2$, and $m=1/2$. In this set of units, the energy
eigenvalues are the minus of the roots of the following algebraic
equation
\begin{equation}\label{root}
\mbox{Ai}\left(x\right)+\frac{1}{2}\alpha^2
x\left[9\mbox{Ai}\left(x\right)+
x\mbox{Ai}'\left(x\right)-\frac{1}{4}x^3\mbox{Ai}\left(x\right)\right]=0.
\end{equation}
So, the energy eigenvalues will be quantized and result in the
following eigenfunctions
\begin{eqnarray}
\psi_n(x)&=&\mbox{Ai}\left(x-E_n\right)+\frac{1}{2}\alpha^2
(x-E_n)\Big[9\mbox{Ai}\left(x-E_n\right)+\nonumber\\&+& (x-E_n)
\mbox{Ai}'\left(x-E_n\right)-\frac{1}{4}(x-E_n)^3\mbox{Ai}\left(x-E_n\right)\Big],
\end{eqnarray}
where $E_n$ should satisfy eq.~(\ref{root}). Figure \ref{fig1} shows
the resulting normalized ground state and first excited state
eigenfunctions for perturbed and unperturbed Hamiltonians with
$\alpha=0.1$. Moreover, we present the first ten energy eigenvalues
for $\alpha=0.01$ in table \ref{tab1}. These results, as we have
expected, show that in the presence of GUP the energy spectrum
slightly increases. So, the assumptions of a minimal length and a
maximal momentum result in a positive shift in the energy spectrum
of a quantum bouncer. However, as table \ref{tab1} shows, this
positive shift for the case of $H=H_0+H_1+H_2$ is smaller with
respect to the case that we relax the assumption of a maximum
momentum $H=H_0+H_2$. This is due to the fact that when we impose an
upper bound on the momentum, we  actually eliminate the contribution
of highly excited states. This effect is also observed in the
perturbation analysis of a particle in a box and the harmonic
oscillator in the GUP formalism \cite{p1}.

We can also use these results for the case of ultra cold neutrons in
an experiment with high precision neutron gravitational spectrometer
which has been demonstrated few years ago \cite{291,292,293}. In
fact, the observation of spontaneous decay of an excited state and
graviton emission in this experiment would be a Planck-scale physics
effect \cite{30}. The transition probability in the quadrupole
approximation and in the presence of GUP is \cite{NP,30}
\begin{eqnarray}
\Gamma^{\mbox{\tiny{GUP}}}_{k\rightarrow
n}\simeq\left(1+\frac{5\Delta \lambda_{kn}}{\lambda_{k}-
\lambda_{n}}\right)\Gamma_{k\rightarrow n},
\end{eqnarray}
where $-\lambda_{n}$ are the zeros of the Airy function,
$\Delta\lambda_{n}=\displaystyle\frac{E_n}{E_0}-\lambda_{n}$,
$E_{0}=mg/\beta$,
$\Delta\lambda_{kn}=\Delta\lambda_{k}-\Delta\lambda_{n}$, and
\begin{eqnarray}
\Gamma_{k\rightarrow n}=\frac{512}{5(\lambda_k-\lambda_n)^3}
\left(\frac{m}{M_{Pl}}\right)^{2}\frac{E^{5}_{0}c}{\beta^{4}(\hbar
c)^{5}}.
\end{eqnarray}
For instance, the probability of the spontaneous graviton emission
from the first excited state to the ground state in the absence of
GUP is $\Gamma_{1\rightarrow 0}\sim10^{-77}\mbox{s}^{-1}$ \cite{30}.
Incorporation of the GUP effect causes a shift in transition rate
that are summarized in the last column of table 1. So it is
essentially possible to find the effects of the generalized
uncertainty principle on the transition rate of neutrons bouncing
above a mirror in the Earth's gravitational field. Now, using table
\ref{tab1}, we can compare the effect of $H_1$ \cite{NP} and
$H_1+H_2$ on the transition probability. Since for all states we
have $\Delta\lambda^{(02)}_{kn}-\Delta\lambda^{(012)}_{kn}>0$, we
find that the existence of both a minimal length and a maximal
momentum reduces the transition rate with respect to the presence of
a minimal length alone.

To show the consistency of our approach with other quantum gravity
models, let us consider the Hamiltonian of a (1+1)-dimensional
quantum gravity model in the post-Newtonian approximation as
$H=H^0+H'$ where \cite{MY}
\begin{equation}\label{9[MY]}
H^0=\frac{p^2}{m}+2\pi Gm^2|r|,
\end{equation}
and
\begin{equation}\label{10[MY]}
H'=-\frac{p^4}{4m^3c^2}+\frac{4\pi G}{c^2}|r|p^2.
\end{equation}
Note that the second part of (\ref{9[MY]}) has the form $V(x)=mgx$
upon choosing $g\rightarrow 2\pi Gm$ and the first part of
(\ref{10[MY]}) has the form of $\alpha^2H_2$ upon choosing
$\alpha^2\rightarrow \displaystyle\frac{1}{20m^2c^2}$.

At this point, let us derive a relation for
$\left|\displaystyle\frac{\Delta E_n}{E_n}\right|$, where $\Delta
E_n=E_n-E_n^0$. By expanding eq.~(\ref{zero}) around the
unperturbed solutions, the first and the last terms in the square
bracket are negligible in comparison with the second term and
using
\begin{equation}
\mbox{Ai}\left(-\frac{\beta}{mg}E_n\right)\simeq-\frac{E_n-E_n^0}{mg}
\mbox{Ai}'\left[\beta\left(x-\frac{E_n}{mg}\right)\right]\Bigg|_{x=0}+\ldots
\end{equation}
we find
\begin{equation}
\left|\frac{\Delta E_n}{E_n}\right|\sim m\alpha^2 E_n,
\end{equation}
which is in agreement with the general result of ref.~\cite{p1}.
Also from eq.~(\ref{zero}) we have $E_n\approx -\beta^{-1}mga_n$
where $a_n$ are the zeros of the Airy function and
$\beta=\left(\frac{2m^2g}{\hbar^2}\right)^{1/3}$. Thus, we find
$\left|\frac{\Delta E_n}{E_n}\right|\sim
m^2\alpha^2g\left(\frac{\hbar^2}{m^2g}\right)^{1/3}(-a_n)$.
Finally, by choosing $g\rightarrow 2\pi Gm$ and
$\alpha^2\rightarrow \displaystyle\frac{1}{20m^2c^2}$ we obtain
\begin{equation}
\left|\frac{\Delta E_n}{E_n}\right|\sim \frac{1}{20}\frac{(2\pi\hbar
G)^{2/3}}{c^2}(-a_n),
\end{equation}
which agrees with the results of ref.~\cite{MY}.

\begin{table}
\begin{center}
\begin{tabular}{|c|c|c|c|c|c|}
\hline
  $n$& $H_0$     & $H_0+H_2$ & $\Delta\lambda^{(02)}_{n(n-1)}$&$H_0+H_1+H_2$ & $\Delta\lambda^{(012)}_{n(n-1)}$ \\\hline
    0& 2.3381    &  2.3392   & -                     & 2.3384       &  -    \\
    1& 4.0879    &  4.0913   & 2.3 $\times$ $10^{-3}$& 4.0888       & 0.6 $\times$ $10^{-3}$\\
    2& 5.5206    &  5.5267   & 2.7 $\times$ $10^{-3}$& 5.5221       & 0.6 $\times$ $10^{-3}$ \\
    3& 6.7867    &  6.7960   & 3.2 $\times$ $10^{-3}$& 6.7891       & 0.9 $\times$ $10^{-3}$ \\
    4& 7.9441    &  7.9569   & 3.5 $\times$ $10^{-3}$& 7.9475       & 1.0 $\times$ $10^{-3}$ \\
    5& 9.0226    &  9.0391   & 3.7 $\times$ $10^{-3}$& 9.0271       & 1.1 $\times$ $10^{-3}$ \\
    6& 10.040    &  10.061   & 4.5 $\times$ $10^{-3}$& 10.046       & 1.5 $\times$ $10^{-3}$ \\
    7& 11.008    &  11.033   & 4.0 $\times$ $10^{-3}$& 11.016       & 2.0 $\times$ $10^{-3}$ \\
    8& 11.936    &  11.965   & 4.0 $\times$ $10^{-3}$& 11.946       & 2.0 $\times$ $10^{-3}$ \\
    9& 12.829    &  12.862   & 4.0 $\times$ $10^{-3}$& 12.841       & 2.0 $\times$ $10^{-3}$ \\ \hline
\end{tabular}
\end{center}
\caption{The first ten quantized energies of a quantum bouncer in GUP formalism for $\hbar=1$, $g=2$, $m=1/2$, and  $\alpha=0.01$.}\label{tab1}
\end{table}

As the final remark, let us estimate the actual magnitude of the
GUP corrections to the quantum systems. To do this end, we need to
use the numerical values of the fundamental constants $c$, $\hbar$
and $G$ and the neutron's mass ($\sim 10^{-27}\,$Kg) and energy
($\sim 10^{-12}\,$eV) \cite{291,292,293} in the calculations
presented above i.e.
\begin{equation}
\left|\frac{\Delta E_n}{E_n}\right|\sim \alpha_0^2
\frac{\ell_{Pl}^2}{\hbar^2} m E_n\sim   10^{-60}\alpha_0^2.
\end{equation}
It is usually assumed that the dimensionless parameter $\alpha_0$
is of the order of unity \cite{main}. In this case, the minimal
measurable length is the Planck length $\ell_{pl}$ (\ref{Lmin}).
Therefore, the $\alpha$-dependent terms are important only when
energies (lengths) are comparable to the Planck energy (length).
So if we assume $\alpha_0\sim1$, the relative change in the
Neutron's energy is of the order of ${\cal O}(10^{-60})$ which as
we have expected is very tiny. However, if we relax this
assumption, since the accuracy of Nesvizhevsky experiments is
about $\displaystyle\frac{\Delta z}{z}\sim
\displaystyle\frac{\Delta E}{E}\sim10\%$ \cite{291,292,293}, where
$\Delta z$ denotes the uncertainty of the Neutron's position and
$E=mgz$, the upper bound of $\alpha_0$ would be
\begin{equation}
\alpha_0\leq10^{29},
\end{equation}
which is weaker than that predicted by the electroweak scale
$\alpha_0\leq10^{17}$ \cite{54,main}. Therefore, the more accurate
measurements indeed reduce the upper bound on $\alpha_0$ or show
the effects of GUP on the spectrum of the ultra cold neutrons.

\section{Conclusions}\label{sec5}
In this paper we have studied the effects of a recently proposed
Generalized Uncertainty Principle on quantum mechanical systems.
This form of GUP is consistent with various candidates of quantum
gravity such as string theory, doubly special relativity and black
hole physics which also implies a maximum observable momentum.  We
showed that the presence of a minimal length and a maximal
momentum results in the modification of all Hamiltonians in
quantum mechanics. In fact, the modified Hamiltonians contain two
additional terms proportional to $\alpha p^3$ and $\alpha^2 p^4$
which result in a fourth-order generalized Schr\"odinger equation.
For the case of a quantum bouncer, to avoid unphysical solutions,
we solved it in the momentum space as a first-order differential
equation and obtained the energy eigenvalues and eigenfunctions up
to the second order of the GUP parameter. We showed that, although
the additional term $H_1$ has no first-order contribution in the
solutions, it has a second-order contribution and reduces the
effect of the second term $H_2$ on the energy spectrum. In other
words, the upper limit on the momentum excludes the contribution
of highly excited states. This result is also in agreement with
previous perturbative studies regarding other quantum mechanical
systems. Moreover, the presence of a maximal momentum reduces the
transition rate of ultra cold neutrons bouncing above a mirror in
the Earth's gravitational field in comparison with the case that
only $H_2$ is present. We note that if these effects be confirmed
by future experiments, they could make a deep influence on our
understanding about our surrounding universe and also on ultimate
formulation of the quantum gravity proposal.

\subsection*{Acknowledgments}The work of Kourosh Nozari is
supported financially by the Research Council of the Islamic Azad
University, Sari Branch, Sari, Iran.

\end{document}